\documentclass[preprintnumbers,superscriptaddress,pra,showkeys]{revtex4}
\usepackage{amsmath}

\input{tcilatex}
\begin{document}

\title{The vanishing of heat capacity as thermodynamic third law implies
existence of singular systems}
\author{S. F. Xiao}
\email{xiaosf@lingnan.edu.cn}
\affiliation{College of Physical Science and Technology, Lingnan Normal University,
Zhanjiang 524048, China}
\affiliation{School for Theoretical Physics, School of Physics and Electronics, Hunan
University, Changsha 410082, China}
\author{Q. H. Liu}
\email{quanhuiliu@gmail.com}
\affiliation{School for Theoretical Physics, School of Physics and Electronics, Hunan
University, Changsha 410082, China}
\date{\today }

\begin{abstract}
A corollary of the third law of thermodynamics is that the heat capacities
of a system approach zero as the temperature approaches absolute zero Kevin.
Many have attempted to take the corollary as the third law, but two
counterexamples has been constructed explicitly. We present a theorem that
the vanishing of heat capacity as the third law implies an existence of
singular systems, and two known counterexamples are illustrations of the
theorem.
\end{abstract}

\maketitle

\section{Introduction}

The attempt to derive the thermodynamic third law from the second plus the
vanishing of the specific heats (VSH) at absolute zero Kevin and the trial
to take the VSH as the third law itself have a long history and continued to
nowadays. \cite{chen1,chen2} Some are strongly against these attempts. \cite%
{comment1,comment2,comment3,comment4} It is evident that none of directly
experimental falsifiable/justifiable quantity such as the specific heat is
qualified to express the fundamental principles of a theory unless it is an
empirical one. \cite{dyson} Usually the third law has three statements:
Nernst theorem states that the isothermal change of entropy $({\Delta }S)_{T}
$ vanishes as the temperature $T$ vanishes, and Planck's statement asserts
that zero-point entropy is zero $S\left( 0\right) =0$, and unattainability
statement says that the absolute zero cannot be reached by any finite
physical means. All these statements are mutually equivalent in physics. We
note that all previous comments \cite{comment1,comment2,comment3,comment4}
overlook a simple fact:\ the VSH as the third law implies an existence of
the singular zero-point entropy for the possible system. In this sense, the
VSH statement is totally different from the standard ones.

Let us start from the definition of the heat capacity $C_{x}$ with some
external parameter $x$ be fixed,%
\begin{equation}
C_{x}\equiv \underset{\Delta T\rightarrow 0}{\lim }\left( \frac{\Delta Q}{%
\Delta T}\right) _{x}=T\left( \frac{\partial S}{\partial T}\right) _{x}.
\label{cx}
\end{equation}%
We have in consequence,%
\begin{equation}
S\left( x,T\right) -S\left( x,T=0\right) =\int_{0}^{T}\frac{C_{x}(T^{\prime
})}{T^{\prime }}dT^{\prime }.  \label{sx}
\end{equation}%
In mathematics, from this equation it is an easy task to find a
solution/system that satisfies two conditions,%
\begin{equation}
\underset{T\rightarrow 0}{\lim }C_{x}(T^{\prime })=0\text{, and }S\left(
x,T=0\right) \text{ }=const.\text{.}  \label{cond1}
\end{equation}%
It is the third law although the constant is frequently zero as $const.=0$.
Thus we have only one choose in the following.

\textbf{THEOREM}. The VSH statement of the third law implies an existence of
thermodynamic systems of singular zero-point entropy, i.e., satisfying
following two conditions simultaneously,%
\begin{equation}
\underset{T\rightarrow 0}{\lim }C_{x}(T^{\prime })=0\text{ and }S\left(
x,T=0\right) \text{ singular.}  \label{cond2}
\end{equation}%
By the singular, we mean divergent or undefinable or free to specify. 

Since the singular zero-point entropy is completely excluded from the
standard statements of the third law, the VSH statement sharply contradicts
to the standard ones. Two such solutions to equation (\ref{sx}) with
conditions (\ref{cond2}), independently given by Mattis \cite{comment3} and
Landsberg \cite{comment1} can be taken as illustrations of this theorem.

Section II and III will discuss the Mattis model and Landsberg model,
respectively. Section IV is the discussion and conclusion.

\section{Mattis model}

Mattis constructed the following theoretical model in his textbook. \cite%
{comment3} Consider a hypothetical system, where the average heat capacity
per particle $c(T)$ is,%
\begin{equation}
c(T)=\frac{c_{0}}{\ln (1+T_{0}/T)},  \label{1}
\end{equation}%
where $c_{0}$ and $T_{0}$ are fixed parameters. One can easily check that $%
c(T)$ meets following requirement,%
\begin{equation}
\underset{T\rightarrow 0}{\lim }c(T)=0.  \label{2}
\end{equation}%
However, the entropy $s(0)$ per particle is singular, as shown in the
following.

Formally, the relation (\ref{sx}) gives following equation between $s$ and $%
c(T)$,%
\begin{equation}
s(T)-s(\varepsilon )=\int_{\varepsilon }^{T}\frac{c(T^{\prime })}{T^{\prime }%
}dT^{\prime }  \label{3}
\end{equation}%
where an infinitesimal quantity $\varepsilon \left( \rightarrow 0\right) $
is introduced to make the singular integral possibly integrable. The third
law of thermodynamics indicates that following two relations hold true,%
\begin{equation}
\underset{T\rightarrow 0}{\lim }s(T)=0\text{ and }\underset{\varepsilon
\rightarrow 0}{\lim }s(\varepsilon )=0.  \label{4}
\end{equation}%
Assume that the ratio $c(T)/T$ is well-behaved as $T\rightarrow 0$, we can
consider that in a small temperature interval $[\varepsilon ,T]$, the
integral (\ref{3}) can be written as,%
\begin{equation}
s(T)-s(\varepsilon )\approx \frac{c(T)}{T}\left( T-\varepsilon \right) =c(T)-%
\frac{c(T)}{T}\varepsilon .  \label{5}
\end{equation}%
The last term is singular, because the order of two limits $\varepsilon
\rightarrow 0$ and $T\rightarrow 0$ cannot be interchanged. First taking the
limit $\varepsilon \rightarrow 0$, it is zero; and first taking $%
T\rightarrow 0$, it is divergent. Thus, zero-point entropy $s(0)$ is not
analytic. To see this fact more clearly, let us first taking the limit $%
\varepsilon \rightarrow 0$, we have,%
\begin{equation}
s(T)\approx c(T).  \label{6}
\end{equation}%
This result cannot reproduce $c(T)$ (\ref{1}) for we have from (\ref{cx}),%
\begin{equation}
c(T)\equiv T\frac{ds(T)}{dT}\rightarrow \frac{T_{0}}{c_{0}\left(
T+T_{0}\right) }\left( \frac{c_{0}}{\ln (1+T_{0}/T)}\right) ^{2}=\frac{T_{0}%
}{c_{0}\left( T+T_{0}\right) }c(T)^{2}.  \label{7}
\end{equation}%
Therefore, the Mattis model has the vanishing heat capacity $\underset{%
T\rightarrow 0}{\lim }c(T)=0$ but does not has a well-defined zero-point
entropy $s(0)$, thus violating the third law of thermodynamics.

\section{Landsberg model}

Landsberg constructed a model based on the ideal gas equation of state, \cite%
{comment1} 
\begin{equation}
pV=NkT.  \label{8}
\end{equation}%
Integration of the following Maxwell relation,%
\begin{equation}
\left( \frac{\partial S}{\partial V}\right) _{T}=\left( \frac{\partial p}{%
\partial T}\right) _{V}=\frac{Nk}{V},  \label{9}
\end{equation}%
gives the entropy with its extensive nature be assumed,%
\begin{equation}
S=Nk\ln \left( \frac{V}{V_{0}N}D(T)\right) ,  \label{10}
\end{equation}%
where $V_{0}$ is a constant with the volume dimension, and $D(T)$ is an
arbitrary dimensionless function of temperature. The heat capacity is,%
\begin{equation}
C_{V}=NkT\frac{d\ln D(T)}{dT}.  \label{11}
\end{equation}%
Given one boundary condition $\underset{T\rightarrow 0}{\lim }C_{V}/Nk=0$,
there are many solutions $D(T)$ that meet the requirement $S_{V}\left(
0\right) /Nk\neq 0$. By the Landsberg model, we mean a peculiar choose of $%
D(T)$, \cite{comment1}%
\begin{equation}
D(T)=\exp (\frac{kT}{2b}),  \label{12}
\end{equation}%
where $b$ is a constant. The heat capacity is thus $C_{V}=Nk(kT/2b)$($=0$
when $T=0$). The entropy is,%
\begin{equation}
S=Nk\ln \left( \frac{V}{V_{0}N}\right) +Nk\frac{kT}{2b}.  \label{13}
\end{equation}%
Violation of the third law of thermodynamics becomes evident for we have
with the ideal gas equation of state,%
\begin{equation}
\underset{T\rightarrow 0}{\lim }\frac{S}{Nk}=\ln \left( \frac{V}{V_{0}N}%
\right) =\ln \left( \frac{kT}{pV_{0}}\right) \rightarrow -\infty ,
\label{14}
\end{equation}%
which holds at any finite value of pressure $p$.

\section{Discussions and Conclusions}

The Mattis model and the Landsberg model appear different, but both do not
have a well-defined zero-point entropy per particle $s(0)$, thus violating
the third law of thermodynamics. Neither Mattis nor Landsberg has noticed
that it is due to the singular zero-point entropy. The theorem presented in
the Introduction section tells that not only these two models but also all
(counter)examples must possess the singularity. In other words, if the VSH
at $T=0$ can be taken as one statement of the third law, we can seek for the
a system that has such a singularity, which is completely excluded from
standard statements of the the third law. In this sense, the VSH statement
is faulty can never be considered equivalent with the existence ones.

We hope that our theorem with illustrations is helpful to terminate the
further attempt to take the VSH at $T=0$ as one form of the third law and
further trial to prove its equivalence to other forms.

\begin{acknowledgments}
This work is financially supported by the Hunan Province Education Reform
Project under Grants No. HNJG-2022-0506 and No. HNJG-2023-0147. QH is
indebted to the members of Online Club Nanothermodynamica (Founded in June
2020), and members of National Association of Thermodynamics and Statistical
Physics Teachers in China, for fruitful discussions.
\end{acknowledgments}

\end{document}